\documentstyle{article}

\title{\bf A Monte Carlo Study of Pairwise Comparison}

\author{Michael W. Herman\\
\and
Waldemar W. Koczkodaj\thanks{partially supported by
the Natural Sciences and Engineering Research Council of Canada (NSERC)
and by the Ministry of Northern Development and Mines
through the Northern Ontario Heritage Fund Corporation} \\
Department of Mathematics and Computer Science \\
Laurentian University, Sudbury, Ontario, Canada P3E 2C6 \\
waldemar@ramsey.cs.laurentian.ca
}

\date{Sudbury, \today}
\begin{document}
\maketitle


\section{Basics of Pairwise Comparison Method}

Making comparative judgments of intangible stimuli or criteria
(e.g. the degree of an environmental hazard or pollution factors)
involves not only imprecise or inexact knowledge but also inconsistency
in our own judgements. The improvement of knowledge elicitation by
controlling the inconsistency of experts' judgments is not only
desirable but absolutely necessary.
Due to space limitations, the reader's familiarity
with \cite{DuKo94} is assumed although this paper is not a continuation
of \cite{DuKo94} but addresses different aspects of the same theory.
Only the essential concepts of the pairwise comparison method
are presented here.

The basic model of knowledge engineering is based on teamwork in
which a knowledge engineer mediates between human experts and the
knowledge base. The knowledge engineer elicits knowledge from the
experts, refines it with the experts, and represents it in the
knowledge base. Arrow's impossibility theorem states that no solution
to the problem of a group ranking exists under general assumptions
(see \cite{Arr51,Fre86}); however, a constructive
algorithm exists under modified (but still practical) assumptions
when we are able to compare the stimuli in pairs.

The pairwise comparison methodology introduced by Thurstone
in 1927 (see \cite{Thur27}) can be used as a powerful inference
tool and knowledge acquisition technique in knowledge-based systems.
Some of the notable applications are related to projects
of national importance, e.g., decisions on nuclear power plants
in Holland (\cite{Nijk90}) and a transportation system in
Sudan(\cite{SaVa82}).

The practical and theoretical virtue of the pairwise comparison
methodology is its simplicity. The goal of pairwise comparisons is
to establish the relative preferences of $n$ stimuli in situations
in which it is impractical (or sometimes even meaningless)
to provide estimates for the stimuli.
To this end, an expert (or a team of experts) provides pairwise
comparison coefficients $a_{ij}>0$, which are
meant to be a substitute for the quotients $s_i/s_j$ of the unknown
(or even undefined) values of the stimuli $s_i, s_j>0$. The
quotients $s_i/s_j$ are also sometimes called {\em relative weights}
in the literature. 

For the sake of our exposition we define an $n \times n$
pairwise comparison matrix simply as a square matrix ${\bf A}=[a_{ij}]$ such
that $a_{ij}>0$ for every $i,j=1, \ldots ,n$. A pairwise comparison
matrix ${\bf A}$ is called {\em reciprocal} if $a_{ij} = \frac{1}{a_{ji}}$
for every $i,j=1, \ldots ,n$ (then
automatically $a_{ii}=1$ for every $i=1, \ldots ,n $). Let


\begin{center}
${\bf A}=\left| \begin{array}{cccc}
                  1 & a_{12} & \cdots & a_{1n} \\
                  \frac{1}{a_{12}} & 1 & \cdots & a_{2n} \\
                  \vdots & \vdots & \vdots &\vdots \\
                  \frac{1}{a_{1n}} & \frac{1}{a_{2n}} & \cdots & 1
                \end{array} \right|$
\end{center}

\noindent where $a_{ij}$ expresses an expert's relative preference of
stimuli $s_i$ over $s_j$.

A pairwise comparison matrix ${\bf A}$ is called consistent if
$a_{ij} \cdot a_{jk}=a_{ik}$ for every $i,j,k=1, \ldots ,n$. While every
consistent matrix is reciprocal, the converse is false in general.

Consistent matrices correspond to the ideal situation in which there
are exact values $s_1, \ldots , s_n$ for the stimuli. The quotients
$a_{ij}=s_i/s_j$ form a consistent matrix. Conversely, the starting
point of the pairwise comparison inference theory is Saaty's theorem
(see \cite{Saaty77}) which states that for every $n \times n$ consistent
matrix ${\bf A}=[a_{ij}]$ there exist positive real numbers $s_1, \ldots s_n$
such that $a_{ij}=s_i/s_j$ for every $i,j=1, \ldots , n$.
The vector $s=[s_1, \ldots s_n]$ is unique up to a multiplicative
constant.

The challenge to the pairwise comparison method comes from
the lack of consistency of the pairwise comparison matrices which
arise in practice (while as a rule, all the pairwise
comparison matrices are reciprocal). Given an $n \times n$ matrix ${\bf A}$
which is not consistent, the theory attempts to provide
a consistent $n \times n$ matrix ${\bf C}$ which differs from matrix ${\bf A}$
``as little as possible''. One of the possible solutions to this problem
was proposed by Saaty (see \cite{Saaty77}). Let $s=[s_1, \ldots ,s_n]$
be the eigenvector of ${\bf A}$ corresponding to $\sigma$, the largest
eigenvalue in modulus of ${\bf A}$. By the Frobenius Theorem (see \cite{Gant59}), the
eigenvalue $\sigma$ is unique, positive, and simple. Furthermore, the
vector $s$ can be chosen with all components positive.

There has been an ongoing discussion about which method for finding solutions
to a pairwise comparison matrix is better (\cite{SaVa84,Craw87}).
Amongst the strongest competitors are the Least Squares ($LS$),
the Logarithmic Least Squares ($LLS$), the Geometric Means ($GM$),
and the Eigenvector ($EV$) methods (See \cite{Jens84} and
\cite{SaVa84} for details). No decisive analytical proof has been
published yet (to our knowledge). Furthermore, it is not clear whether
or not an analytical proof can be devised (this issue is addressed in
the conclusions of this paper). This seems to be a good reason for
formulating the problem as an empirical experiment using
a Monte Carlo approach.

Only two of the above methods will be considered in this paper,
because of their practical importance:
\begin{itemize}
\item $GM$ for simplicity (all one needs is a pocket calculator in most
cases),
\item $EV$ for its claimed superiority (\cite{SaVa84}) and its
mathematical elegance which provides useful interpretations of
the resulting weights.
\end{itemize}

We need to consider reciprocal matrices of orders 4 to 7.
There is a proof that solutions to $R3$ ($Rn$ will stand for
a reciprocal matrix of order n) by $GM$ and $EV$ are exactly the same;
and orders above seven are impractical due to the large number
of comparisons required.
There are two important problems to examine:
\begin{itemize}
\item How to measure the accuracy of solutions?
\item How to generate, let us say, $``not-so-inconsistent''$ ($NSI$ for short)
reciprocal matrices?
\end{itemize}

\section{Comparison of solutions}

Having a solution (obtained by any of the above methods) we can
reconstruct the reciprocal matrix (by simple divisions of appropriate 
components of the solution vector $s$). The distance (e.g., Euclidean
or Tchebychev) between the given and reconstructed matrices will
be used as a measure of the accuracy of the solution.

Solutions of matrices are obtained by $GM$ (geometric means, see
\cite{Craw87}) and $EV$ (eigenvector method, see \cite{Saaty77})
and normalized to one (by dividing each component by the sum of
all components). There is no way, however, of saying which solution
(obtained by $GM$ or $EV$) is better just by looking at them.
For each vector of weights (a solution), $s_i, i=1,\ldots,n,$ where all $s_i$
are positive and $s_i \in [1/max, max]$ where $max$ is a scale
maximum, we reconstruct the matrix ${\bf A}=[a_{ij}]$ where $a_{ij}=s_i/s_j$
for $i,j=1,\ldots,n$. The distance between the matrix ${\bf A}$ and the original
matrix is considered to be a measure of the solution accuracy.
For a given fully consistent matrix this distance is equal to 0
(as expected), but for an inconsistent matrix it is greater than 0.

Two metrics have been applied:
\begin{itemize}
\item modified Euclidean, $\frac{\sqrt{\sum_{i,j=1}^n (a_{ij}-b_{ij})^2}}{n^2}$,
  (note the division by the total number of elements)
\item Tchebychev (also known as the infinity metric) , $\max(|a_{ij}-b_{ij}|)$
for $i,j=1,\ldots,n$.
\end{itemize}

The division of the Euclidean distance by the number of matrix elements
is necessary for comparing it with the Tchebychev distance which is the
maximum distance between two corresponding matrix elements. Needless
to say, this operation is just a simple scaling and has no influence on the
final results.

No formal theorem has been proven (to our knowledge) showing that for
a given matrix {\bf A} which is close enough to a consistent matrix {\bf C},
the solutions of {\bf C} are acceptable approximations for the problem
described by {\bf A}.
The inference that for a matrix with small inconsistency, the solutions
of weights are close enough to real values of stimuli is purely speculative.

\section{Generating $NSI$ matrices}

One should note that in itself the randomization of matrices is a
difficult process but it does not matter if it is not absolutely
perfect. From a statistical point of view, however, it is
imperative that an identical set of randomly generated matrices is used
for both methods since the main goal of this study is to compare
the two methods. Any reasonably random set of matrices should suffice.

The problem of generating consistent reciprocal matrices in a random
way is more complicated than it seems to be. The straight forward method of
generating random matrices, checking their consistency factor ($cf$), and
discarding those which are outside an assumed $cf$ range is impractical.
Our preliminary experiments have established that too many matrices
(of the order of $10^9$ or more) would need to be examined to find
at least one matrix within an assumed range of inconsistencies.
A more constructive approach needs to be implemented. A fully consistent
reciprocal matrix is generated randomly. This matrix is then again randomized
by multiplying the elements above the main diagonal
by a randomizing multiplier ($RM$). The randomizing multiplier $RM$
is constructed by adding or subtracting (at random) a randomly generated
number $\rho \in [0,1]$ which is multiplied by a given deviation ($D$):
\begin{center} $RM=1 \pm \rho \cdot D$ \end{center}
For $D=0$ we get fully consistent matrices.
It is expected that the consistency factor should increase with
increasing $D$. Two consistency factors are used in our analysis:
eigenvalue-based (as introduced in \cite{Saaty77}) and triad-based (as
introduced in \cite{Kocz93}).

The definition of consistency of a pairwise comparison matrix ${\bf A}$,
based on eigenvalues, was introduced by Saaty \cite{Saaty77}. His
consistency definition is given by the following formula:

\begin{center}
$ cf = \frac{\lambda_A - order(A)}{(order(A)-1) \lambda_{random}} $ \\
\end{center}

\noindent where $\lambda_A$ is the largest eigenvalue of the reciprocal 
matrix ${\bf A}$ and $\lambda_{random}$ is the largest eigenvalue of randomly
generated reciprocal matrices of the same order as matrix ${\bf A}$
(see \cite{Saaty77}). For a given order, $\lambda_{random}$ is constant
and many researchers do not use it. This factor is also omitted in our formula
for $cf$ but one may always divide our results by $\lambda_{random}$
(published in \cite{Saaty77}) to get the strict definition of Saaty's $cf$.

It is worthwhile to note that we are unable to establish an explicit
analytical relationship between $cf$ and $D$. In particular we do not know
which $D$ is necessary for a given $cf$ (for either of the definitions).
For each matrix generated for a chosen deviation $D$, we compute
$cf$ (according to each of the above mentioned definitions)
and store it with the other results for further analysis.

\section{Interpretation of results}

In essence, solutions obtained by both methods, geometric means ($GM$) and
eigenvector ($EV$), turned out to be close enough to the given input matrix.
In this respect both methods are accurate enough for most practical
applications. As the enclosed Table 1 demonstrates, the biggest
difference between average deviations of $GM$ and $EV$ solutions is 0.00019
for the Euclidean metric and 0.00355 for the Tchebychev metric.
For practical applications, this precision is far better than expected.
After all we are talking, in most cases, about relative
subjective comparisons and one tenth of a percent is usually below
our threshold of perception.

\vspace{8pt}
\begin{center}
\begin{tabular}{||c||c|c|c|c|c|c|c|c|c||}
\hline
 &
 & \multicolumn{2}{|c|}{$cf$}
 & \multicolumn{3}{|c|}{Euclidean metric}
 & \multicolumn{3}{|c||}{Tchebychev metric} \\ \cline{3-10}
 $Ord$ & $D$ & $triad$ & \quad $\lambda$\ \quad & $dist$ & $diff$ & $wins^1$
                               & $dist$ & $diff$ & $wins^2$
\\ \hline\hline
  &0.1&0.129&0.001&0.0153&0.00000&51.0\%&0.1732&0.00002&56.5\% \\ \cline{2-10}
  &0.2&0.238&0.003&0.0302&0.00001&52.3\%&0.3420&0.00012&59.4\% \\ \cline{2-10}
4 &0.3&0.333&0.007&0.0451&0.00003&53.4\%&0.5080&0.00037&61.1\% \\ \cline{2-10}
  &0.4&0.416&0.014&0.0601&0.00008&54.6\%&0.6744&0.00083&62.1\% \\ \cline{2-10}
  &0.5&0.490&0.022&0.0755&0.00017&55.6\%&0.8444&0.00149&62.7\% \\ \hline
  &0.1&0.160&0.001&0.0138&0.00000&51.0\%&0.2212&0.00003&52.7\% \\ \cline{2-10}
  &0.2&0.293&0.004&0.0273&0.00001&52.2\%&0.4356&0.00020&54.4\% \\ \cline{2-10}
5 &0.3&0.406&0.009&0.0407&0.00003&53.5\%&0.6472&0.00061&55.7\% \\ \cline{2-10}
  &0.4&0.502&0.017&0.0544&0.00009&55.0\%&0.8592&0.00131&56.6\% \\ \cline{2-10}
  &0.5&0.585&0.027&0.0685&0.00019&56.8\%&1.0761&0.00236&57.2\% \\ \hline
  &0.1&0.180&0.001&0.0122&0.00000&51.1\%&0.2546&0.00003&51.5\% \\ \cline{2-10}
  &0.2&0.328&0.004&0.0241&0.00001&52.4\%&0.5017&0.00025&52.7\% \\ \cline{2-10}
6 &0.3&0.450&0.010&0.0361&0.00003&53.9\%&0.7441&0.00082&53.8\% \\ \cline{2-10}
  &0.4&0.553&0.019&0.0483&0.00007&55.7\%&0.9884&0.00173&54.5\% \\ \cline{2-10}
  &0.5&0.640&0.030&0.0610&0.00017&57.8\%&1.2397&0.00300&55.1\% \\ \hline
  &0.1&0.194&0.001&0.0108&0.00000&51.1\%&0.2798&0.00004&51.1\% \\ \cline{2-10}
  &0.2&0.351&0.005&0.0214&0.00001&52.5\%&0.5505&0.00030&52.1\% \\ \cline{2-10}
7 &0.3&0.480&0.011&0.0321&0.00002&54.2\%&0.8172&0.00097&52.8\% \\ \cline{2-10}
  &0.4&0.587&0.020&0.0430&0.00006&56.2\%&1.0850&0.00210&53.6\% \\ \cline{2-10}
  &0.5&0.675&0.033&0.0544&0.00014&58.9\%&1.3624&0.00355&54.2\% \\ \hline
\end{tabular}
\\ \vspace{8pt}
Table 1. Monte Carlo results for Euclidean and Tchebychev metrics
(1,000,000 matrices were tested for each case of order and deviation) \\
\end{center}
\vspace{8pt}

\begin{tabular}{p{0.25in} p{3.75in}}
\multicolumn{2}{l}{Notations and abbreviations used in Table 1:} \\
         &              \\
$Ord$    & matrix order \\
$D$      & deviation to perturb matrix elements \\
$cf$     & consistency factors \\
$triad$  & triad consistency factor \\
$\lambda$& eigenvalue-based consistency factor (computed without
            $\lambda_{random}$ factor) \\
$dist$   & distance between the reconstructed and given matrices \\
$wins^1$ & frequency of winning by $GM$ over $EV$ method in $\%$ \\
$wins^2$ & frequency of winning by $EV$ over $GM$ method in $\%$ \\
$diff$   & difference in accuracy between the winning method
           and the other method \\
\end{tabular}
\\ \vspace{8pt}

The implementation in C was fast. For each deviation $D$, 1,000,000
cases of $NSI$ matrices were generated and analyzed in approximately
10 minutes on a personal computer with a 100MHz Pentium CPU.
The results were checked against those produced by a prototype program
written in APL2. The built-in operations on entire arrays and the operator
``each'' that applies a function itemwise to a vector of arrays permitted a
simplified implementation that was not only shorter, but also easier to
understand and modify. In fact, errors in the C code could easily have gone 
undetected were it not for the availability of the prototype.

Although execution of the APL2 program was slower, fewer matrices were
needed in the calculations used for checking purposes. No attempt was made 
at optimising the APL2 program for speed since it was decided early on that 
it would serve primarily as a reference version on which different ideas 
could be tried easily and changes made quickly, whereas  the C version would 
be used for the full calculations.

The values of the deviation $D$ used were $0.1$, $0.2$, $0.3$, $0.4$, and $0.5$.
It is worthwhile noting that $D=0.5$ may, in extreme cases, result in a
matrix element that is 50 to 150 percent of the original element's value
(that is a ratio of 1:3). The smallest deviation $D=0.1$ perturbs
an element by at most 10\% which creates very small inconsistencies
and has no practical influence on accuracy.

Table 1 also shows that the results generated by the $GM$ method are better
when the Euclidean metric is used while the $EV$ method results are better
for the Tchebychev matric! This should not be a surprise. After all,
geometric means ($GM$) are ``means'', and therefore the Euclidean metric
(which ``equalizes'' differences of all elements) generates better
results for $GM$ than for $EV$ solutions.
On the other hand, the eigenvalue method ($EV$) finds the matrix with
elements as close as possible to the given matrix which
is reflected by the Tchebychev metric (the maximum of the absolute
differences).

The programs for computing all the results are available to interested
readers by email.

\section{Conclusions}

The lack of a clear cut answer as to which method generates more accurate
solutions was a surprise, but the results are nevertheless of practical
importance. For some applications one of methods may be better than the other.
Simplicity and balanced distribution of estimations of all stimuli
would dictate the use of the geometric means method while the more
sophisticated computation of eigenvectors may contribute to a better
precision on each individual component of the solution.

It is improbable that an analytical solution can be devised
in a situation where the results favour $GM$ over $EV$ for one metric
while favouring $EV$ over $GM$ for another metric. 
More importantly, statistical evidence of convergence to a solution
has been observed. The theory of reciprocal matrices states that solutions
for both methods ($GM$ and $EV$) only have sense for fully consistent
matrices. For inconsistent matrices no solution exists. In fact it cannot
exist even if there are only three inconsistent judgements (that is one
triad of inconsistent judgments; for details see \cite{DuKo94}) since having
a solution, we can reconstruct the relative judgments by simple division.
This contradicts the assumption that they were inconsistent.
Until now it has been tacitly assumed that for $``not-so-inconsistent''$
matrices the obtained solutions should be close enough approximations
but (to our knowledge) no formal theorem to this effect has been
published. \\

Table 1 shows that the solutions for more consistent matrices
are more precise (for either of the methods).
This is not surprising, yet not a trivial observation, since a phenomena
of rank reversal exists for extremely inconsistent matrices
(see \cite{SaVa84}). Rank reversal takes place where a solution by one
method ranks stimuli differently from another method. A formal proof
that ``for a matrix with small enough inconsistency no rank reversal
should take place'' still challenges us.

\end{document}